\documentclass{moriond}
\usepackage{heppennames2}

\newcommand{\ttbar}{\PQt{}\PAQt{}\xspace}

\newcommand{\ttH}{\ensuremath{\PQt{}\PAQt{}\PH}\xspace}

\newcommand{\ttll}{\ensuremath{\PQt{}\PAQt{}\ell\ell}\xspace}
\newcommand{\ttlnu}{\ensuremath{\PQt{}\PAQt{}\ell\nu}\xspace}

\newcommand{\tttt}{\ensuremath{\PQt{}\PAQt{}\PQt{}\PAQt}\xspace}
\newcommand{\tHq}{\ensuremath{\PQt{}\PH{}\PQq}\xspace}

\newcommand{\tZq}{\ensuremath{\PQt{}\PZ{}\PQq}\xspace}
\newcommand{\tllq}{\ensuremath{\PQt{}\ell\ell{}\PQq}\xspace}

\newlength\cmsFigWidth
\usepackage{graphicx}

\usepackage{multicol}

\newcommand{\twolss}{\ensuremath{2\ell \textrm{ss}}\xspace}
\newcommand{\threel}{\ensuremath{3\ell}\xspace}

\newcommand*{\eftOp}[4]{\ensuremath{%
		{#4%
			\ifx\empty#3\empty\ifx\empty#1\empty\else^{#1}\fi\else^{#1(#3)}\fi%
			\ifx\empty#2\empty\else_{#2}\fi}%
}}

\newcommand{\ctp}  {\eftOp{}{\PQt\varphi}{}{c}\xspace}

\newcommand{\cpQM} {\eftOp{-}{\varphi Q}{}{c}\xspace}
\newcommand{\cpQa} {\eftOp{3}{\varphi Q}{}{c}\xspace} 
\newcommand{\cpt}  {\eftOp{}{\varphi\PQt}{}{c}\xspace}
\newcommand{\cptb} {\eftOp{}{\varphi\PQt\PQb}{}{c}\xspace}

\newcommand{\ctW}  {\eftOp{}{\PQt\PW}{}{c}\xspace}

\newcommand{\ctZ}  {\eftOp{}{\PQt\PZ}{}{c}\xspace}

\newcommand{\cbW}  {\eftOp{}{\PQb\PW}{}{c}\xspace}

\newcommand{\ctG}  {\eftOp{}{\PQt G}{}{c}\xspace}

\newcommand{\cQla} {\eftOp{3}{Q\ell}{\ell}{c}\xspace}
\newcommand{\cQlM} {\eftOp{-}{Q\ell}{\ell}{c}\xspace}
\newcommand{\cQe}  {\eftOp{}{Q\Pe}{\ell}{c}\xspace}
\newcommand{\ctl}  {\eftOp{}{\PQt \ell}{\ell}{c}\xspace}
\newcommand{\cte}  {\eftOp{}{\PQt \Pe}{\ell}{c}\xspace}
\newcommand{\ctlS} {\eftOp{S}{\PQt}{\ell}{c}\xspace}

\newcommand{\ctlT} {\eftOp{T}{\PQt}{\ell}{c}\xspace}

\newcommand{\PQQ}{\ensuremath{\mathrm{Q}}\xspace}

\newcommand{\GeV}{\ensuremath{\mathrm{GeV}}\xspace}
\newcommand{\cttOne}   {\eftOp{1}{\PQt\PQt}{}{c}\xspace}
\newcommand{\cQQOne}   {\eftOp{1}{\PQQ\PQQ}{}{c}\xspace}
\newcommand{\cQtOne}   {\eftOp{1}{\PQQ\PQt}{}{c}\xspace}
\newcommand{\cQtEight} {\eftOp{8}{\PQQ\PQt}{}{c}\xspace}

\newcommand{\cQqOneThree}   {\eftOp{31}{\PQQ\PQq}{}{c}\xspace}
\newcommand{\cQqEightThree} {\eftOp{38}{\PQQ\PQq}{}{c}\xspace}
\newcommand{\cQqOneOne}     {\eftOp{11}{\PQQ\PQq}{}{c}\xspace}
\newcommand{\ctqOne}        {\eftOp{1}{\PQt\PQq}{}{c}\xspace}
\newcommand{\cQqEightOne}   {\eftOp{18}{\PQQ\PQq}{}{c}\xspace}
\newcommand{\ctqEight}      {\eftOp{8}{\PQt\PQq}{}{c}\xspace}
\newcommand{\pt}{\ensuremath{p_{\mathrm{T}}}\xspace}

\newcommand{\ptljz}{\ensuremath{{\pt}\mathrm{(\ell j 0)}}\xspace}
\newcommand{\ptZ}{\ensuremath{{\pt}\mathrm{(\PZ)}}\xspace}

\newcommand{\Photo}{\includegraphics[height=35mm]{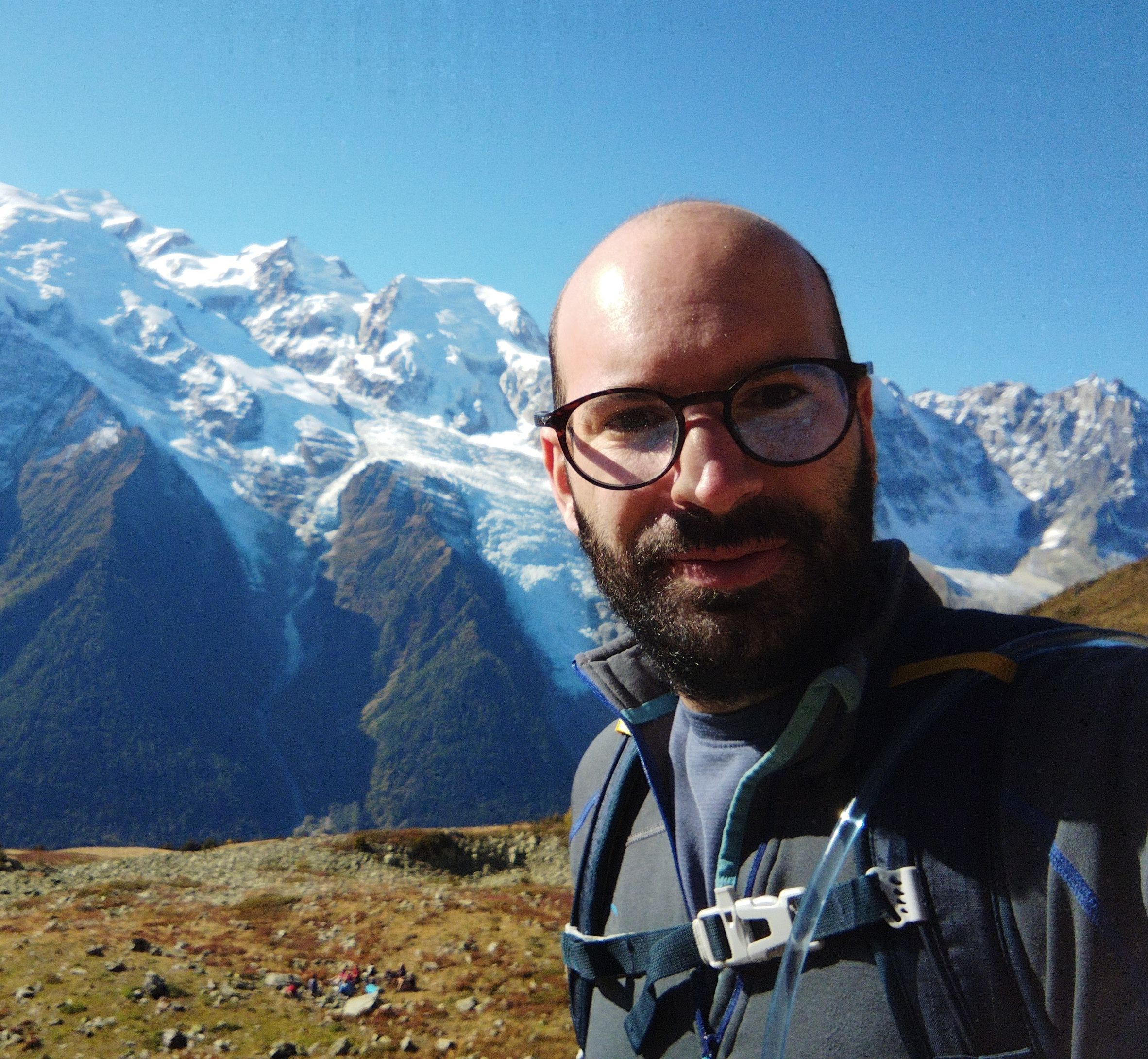}}

\begin{document}
\vspace*{4cm}
\title{Search for EFT in associated top production}

\author{ Sergio Sanchez Cruz on behalf of the CMS Collaboration }

\address{Physik-Institut \\ Universit\"at Z\"urich
	Winterthurerstrasse 190,\\
	8057 Z\"urich, Switzerland}

\maketitle\abstracts{
 A search for new physics in top quark production with additional final-state leptons is performed with 138 $\mathrm{fb}^{-1}$ of proton-proton collision data at $\sqrt{s}$ = 13 TeV. We use the framework of effective field theory to parametrize potential new physics effects in terms of 26 dimension-six EFT operators. The data are divided into several categories based on lepton multiplicity, total lepton charge, jet multiplicities, and b-tagged jet multiplicities. Kinematic variables are used to extract limits simultaneously on the associated set of Wilson coefficients. }

\section{Introduction}

We present a search for physics beyond the Standard Model (BSM) in the framework of Effective Field Theories (EFTs) using a dataset of proton-proton collisions at $\sqrt{s}=13$ TeV collected by the CMS Collaboration~\cite{cms} between 2016 and 2018, which corresponds to a total integrated luminosity of 138 $\mathrm{fb}^{-1}$. The measurement~\cite{TOP22006}, uses different modes of top production in association with additional leptons and supersedes previous results by the CMS in the same channel~\cite{TOP19001} performed using a partial dataset. This updated measurement constrains more EFT operators with an improved analysis strategy and using the largest dataset available.

EFT searches are most useful in a scenario in which the true BSM theory lays just above the experimentally reachable energy scale. In those cases, while BSM particles cannot be directly produced, they generate new interactions that may be observable at lower energy. These interactions can be parametrized as additional operators $\mathcal{O}_j$ to the Standard Model (SM) Lagrangian as if it were an expansion in terms of inverse powers of the ultraviolet (UV) energy scale $\Lambda$,

\begin{equation}
\mathcal{L}_{\mathrm{eff}} = \mathcal{L}_{\mathrm{SM}} + \sum_{j} \frac{c_j}{\Lambda^2} \mathcal{O}_j,
\label{eq:lagrangian}
\end{equation}
where $c_j$ are the Wilson coefficients (WCs) that regulate the strength of the new interactions.

Aiming to be agnostic to the nature of the ultraviolet (UV) theory, we follow a bottom-up approach, looking for deviations that may provide hints on its characteristics and energy scale, interpreting these potential deviations as non-zero values of the WCs.  We consider 26 independent operators simultaneously, described in Tab.~\ref{tab:wc_lst} following the LHC Top WG conventions~\cite{Aguilar_Saavedra:2018ksv}. We have chosen a set of operators  that affects significantly the cross section or kinematics of the signal processes populating the phase space under study (\ttH, \ttll, \ttlnu, \tllq, \tHq and \tttt).

These processes, which typically involve interactions playing a prominent role in several BSM scenarios, have a clear experimental signature giving raise to events with several leptons in the final state. While each of these processes has been studied individually~\cite{HIG19008,HIG21006,HIG18009,TOP21011,TOP18009,TOP20010,TOP21005}, we cannot simply reinterpret the individual measurements in terms of the WCs for two reasons. First, since several of these processes have similar kinematic properties, such measurements are performed in similar and overlapping phase spaces, complicating a statistical combination of the measurement. Secondly, each of the measurements considers the rest of the processes of interest as backgrounds that have the properties predicted by the SM. Instead, any UV theory could modify the properties of several of these processes simultaneously, invalidating this procedure.

Instead, we choose to constrain the WCs by evaluating their effect on the number of expected events for all signal processes using Monte Carlo simulations. This approach, besides overcoming the limitations described above,  allows accounting for EFT effects in signal acceptance and in a fully differential way.

\begin{table}
	\centering
	\caption{List of WCs included in this analysis. }
	\begin{tabular}{ll}
		\hline
		Operator category & WCs \\
		\hline
		Two heavy quarks                   \rule{0pt}{2.8ex} & \ctp, \cpQM, \cpQa, \cpt, \cptb, \ctW, \ctZ, \cbW, \ctG \\
		Two heavy quarks two leptons        \rule{0pt}{2.8ex} & \cQla, \cQlM, \cQe, \ctl, \cte, \ctlS, \ctlT \\
		Two light quarks two heavy quarks  \rule{0pt}{2.8ex} & \cQqOneThree, \cQqEightThree, \cQqOneOne, \cQqEightOne, \ctqOne, \ctqEight\\
		Four heavy quarks                   \rule{0pt}{2.8ex} & \cQQOne, \cQtOne, \cQtEight, \cttOne\\
		\hline
	\end{tabular}
	\label{tab:wc_lst}
\end{table}

\section{Event selection and classification}

The event selection is designed to build phase space regions where the BSM effects are enhanced and can be characterized in terms of the WCs we consider. Firstly, we select events with at least two same-sign or three leptons (electrons or muons), jets, and \PQb-tagged jets in the final state. We make use of the dedicated lepton selection developed in~\cite{HIG18009} to discriminate signal leptons coming from W and Z boson decays from those coming from different origins.
The topology with an opposite-sign lepton pair is typically dominated by \ttbar production and may be covered in dedicated searches~\cite{TOP18006}.  Selected events are then classified based on their topology, building regions that are enriched in the different signal processes, as described in Tab.~\ref{tab:Categories}. 

In the \twolss region we build regions enriched (depleted) in \tttt production by selecting events with three or more (less than three) \PQb-tagged jets. Events are further classified on the sum of the lepton charge, to profit from the charge asymmetry of the $\ttlnu$ signal. Events with exactly three leptons are classified between those with a dilepton pair candidate compatible with a \PZ boson (on-\PZ) and those without such a candidate (off-\PZ). On-\PZ regions are enriched in $\PQt\PAQt(Z\to\ell\ell)$, while off-\PZ regions receive contributions from non-resonant $\ttll$ and $\ttlnu$, hence the latter are split per sum of the lepton charge. Events in all categories are further categorized according to their jet multiplicity.

\begin{table}[h]
	\caption{Event selection and classification into the different signal regions.}
	\label{tab:Categories}
	\centering
	\resizebox{\textwidth}{!}{
		\begin{tabular}{lcccccc}
			\hline
			Event category             & Leptons  & $m_{\ell\ell}$ & b tags & Lepton charge sum & Jets & Kinematic variable \\
			\hline
			2$\ell$ss 2b         & 2        & No requirement                         & 2       & $>0$, $<$0 & 4,5,6,$\ge$7            & \ptljz \\
			2$\ell$ss 3b         & 2        & No requirement                         & $\ge 3$ & $>0$, $<$0 & 4,5,6,$\ge$7            & \ptljz \\
			3$\ell$ off-Z 1b     & 3        &  $\left| m_{\PZ}-m_{\ell\ell}\right|  > 10\GeV$    & 1       & $>0$, $<$0 & 2,3,4,$\ge$5            & $\ptljz$ \\
			3$\ell$ off-Z 2b     & 3        & $\left| m_{\PZ}-m_{\ell\ell}\right| > 10\GeV$    & $\ge 2$ & $>0$, $<$0 & 2,3,4,$\ge$5            & \ptljz \\
			3$\ell$ on-Z  1b     & 3        & $ \left| m_{\PZ}-m_{\ell\ell}\right| \leq 10\GeV$ & 1       & No requirement & 2,3,4,$\ge$5 & \ptZ \\
			3$\ell$ on-Z  2b     & 3        & $\left| m_{\PZ}-m_{\ell\ell}\right| \leq 10\GeV$ & $\ge 2$ & No requirement & 2,3,4,$\ge$5 & \ptZ or \ptljz\\
			4$\ell$              & $\ge$4   & No requirement                         & $\ge 2$ & No requirement & 2,3,$\ge$4   & \ptljz \\
			\hline
	\end{tabular}}
\end{table}

Finally, events are classified based on variables associated with the kinematic properties of the event, allowing us to profit from the fact that BSM contributions may populate a phase space different from the SM ones. By construction, we cannot optimize the choice of this variable optimally for all operators simultaneously, however, we found a good compromise by pursuing the following approach. 

All the categories in the 3$\ell$ on-\PZ region contribute dominantly to the sensitivity to the \ctZ coefficient, except for the 3$\ell$ on-\PZ 2b region with 2 and 3 jets, which contributes dominantly to the sensitivity to \cQqOneThree and \cQqEightThree through their effect on $\PQt\PZ+\PQb$ production. In the former categories, we classify events according to the \pt of the \PZ boson candidate. In the remaining categories, we classify events using the \ptljz variable. This variable is constructed by considering the vectorial sum of the momenta of all pairs of jets and leptons (pairs of jets, pairs of leptons, and pairs formed with a jet and a lepton) in the event, keeping the pair with the largest \pt. 

\section{Signal and background modeling}

As described above, since discriminating kinematically several of the signal processes is not possible, we estimate the contribution from the signal processes under any BSM scenario using Monte Carlo simulations. From equation~\eqref{eq:lagrangian}, we conclude that the amplitude for any process will receive SM contributions in addition to BSM contributions proportional to the corresponding WC. Since the cross section and the yields are proportional to the amplitude squared, for any category, we can write the number of expected events as 

\begin{equation}
n_i = n_{\mathrm{SM},i} + \sum_j l_{i,j} c_j + \sum_{jk} q_{i,jk} c_j c_k,
\label{eq:yields}
\end{equation}

where the first term corresponds to the SM contribution, the second to the interference between the BSM and SM contributions and the third is the pure-BSM contribution and the sums run over the set of WCs. Each of these contributions is estimated with a single Monte Carlo set of samples, generated assuming non-zero values of the WCs to ensure they populate the full phase space, and EFT effects are estimated using the reweighting techniques.~\cite{reweighting}. 

We consider leading order (LO) samples for all the samples, using the dim6top model~\cite{Aguilar_Saavedra:2018ksv} to incorporate the EFT contributions. For \ttH, \ttll, and \ttlnu production we include an additional parton in the final state computed in the matrix element using the MLM scheme. In addition to improving the modeling of the different signal processes at large jet multiplicity, such diagrams introduce new production modes significantly impacting the dependence of the WCs on these processes. 

Even if most processes in our selection are accounted for as signals susceptible to being affected by BSM effects, other processes have a residual contribution to the signal regions. These processes are not significantly affected by the operators that we study and are assumed not to receive sizable BSM contributions. The dominant background in the three- and four-leptons regions is WZ and ZZ production, respectively. We estimate these processes using MC simulations that we validate in dedicated control regions. Additional contributions to all the regions may stem from events containing misidentified leptons or electrons whose charge has been mismeasured and are estimated from dedicated control regions in data, following the procedure described in~\cite{HIG18009}.

\section{Statistical analysis and results}

 We use the likelihood fit to set limits on the WCs. The likelihood considered is built as 

\begin{equation}
	\mathcal{L} = \prod_{i} P(n_i\,| \nu_i(c, \theta)) \prod_{j}  p(\theta_j | \tilde{\theta_j}),
\end{equation}

where $n_i$, $\nu_i$ are the number of observed and expected events in each category,  respectively. The latter is a function of the WCs, $c$, and the nuisance parameters, $\theta$, that parametrize the effect of the systematic uncertainties. $P$ is the probability mass function of a Poissonian distribution and $p$ is the prior pdf for the nuisance parameters. $\nu_i$ is a quadratic function of the WCs as shown by equation~\ref{eq:yields}.

%

\begin{figure}
	\includegraphics[width=\textwidth]{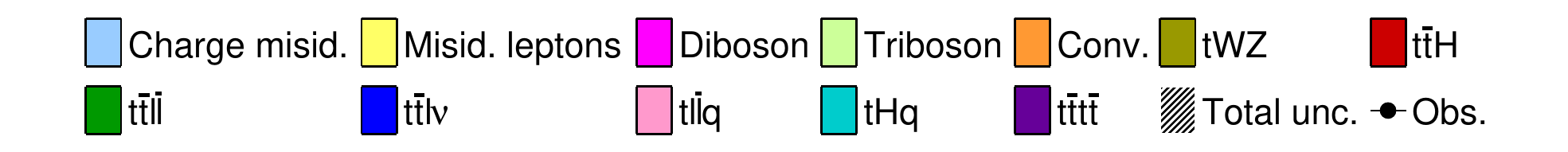}
	\includegraphics[width=\textwidth]{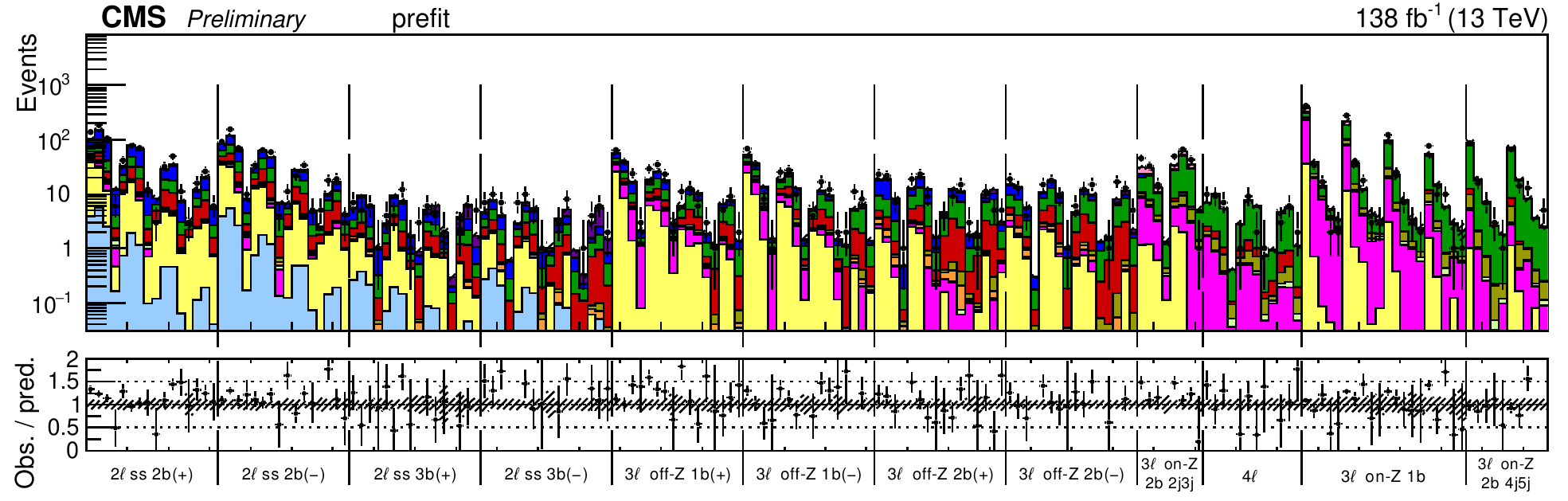}
	\includegraphics[width=\textwidth]{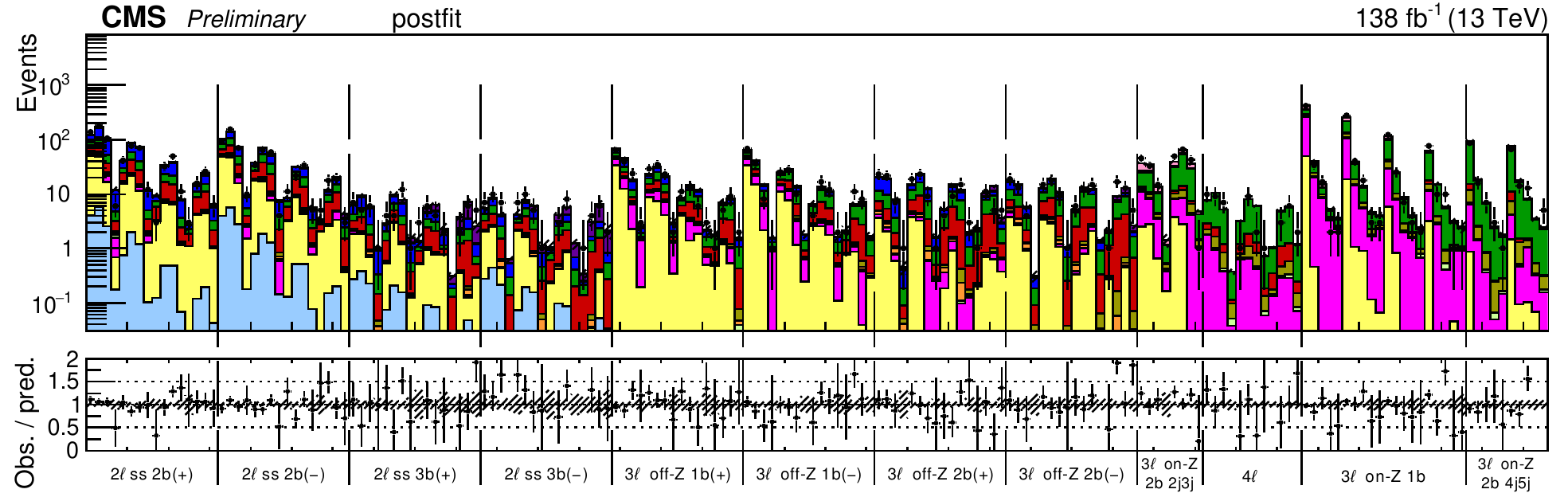}
	\caption{Number of events in the fitted regions compare with the estimation of the signal and background before (top) and after (bottom) the fit is performed.}
	\label{fig:yields}
\end{figure}

The number of observed events in each category is shown in Fig.~\ref{fig:yields} before and after a maximum likelihood fit has been performed. A good agreement is seen of data with the pre- and post-diction, showing good compatibility of data with the fitted model and with the SM. 	

\begin{figure}
	\begin{minipage}{0.55\textwidth}
		\includegraphics[width=\textwidth]{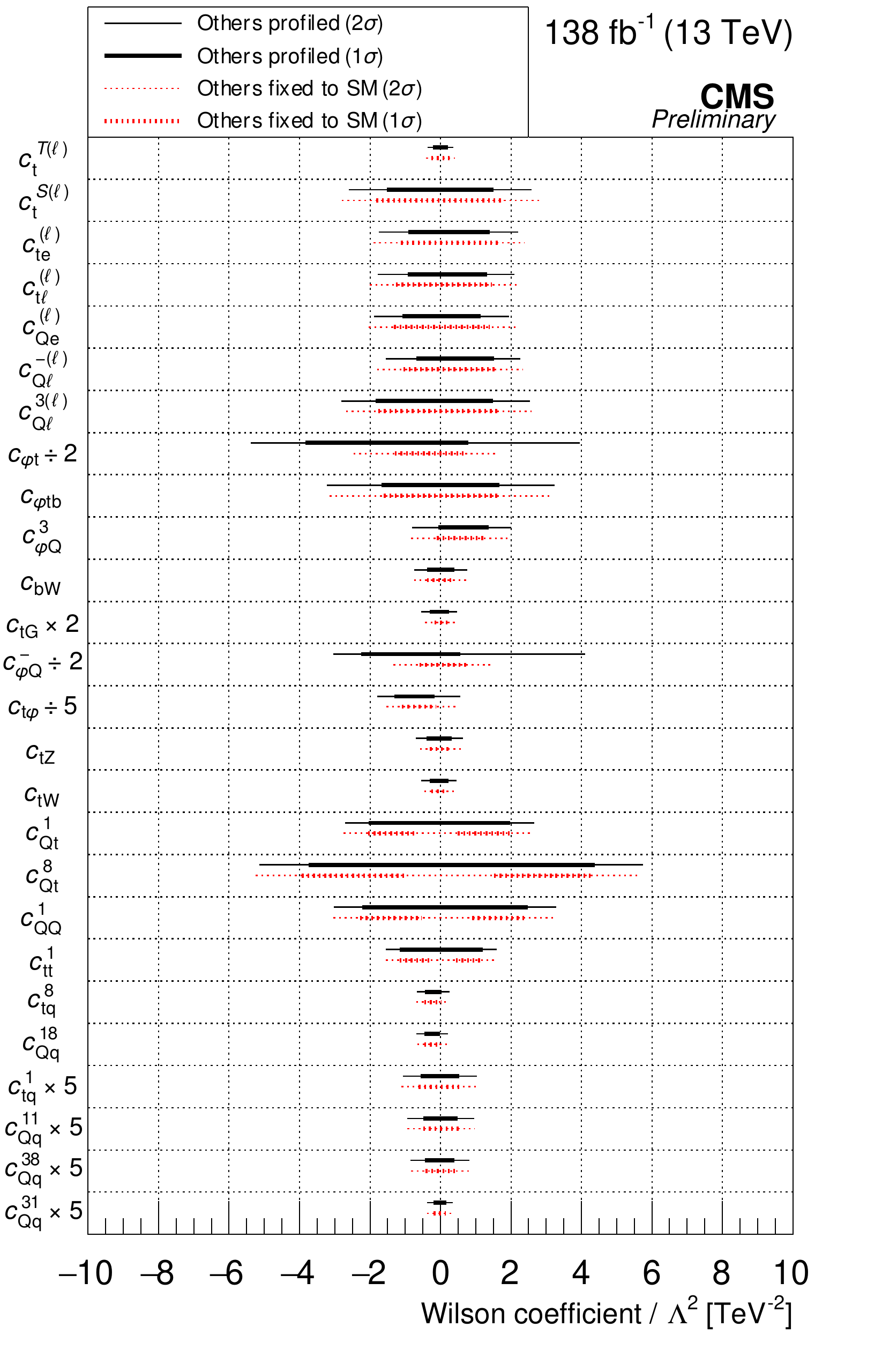}
	\end{minipage} 
	\begin{minipage}{0.45\textwidth}
		\includegraphics[width=\textwidth]{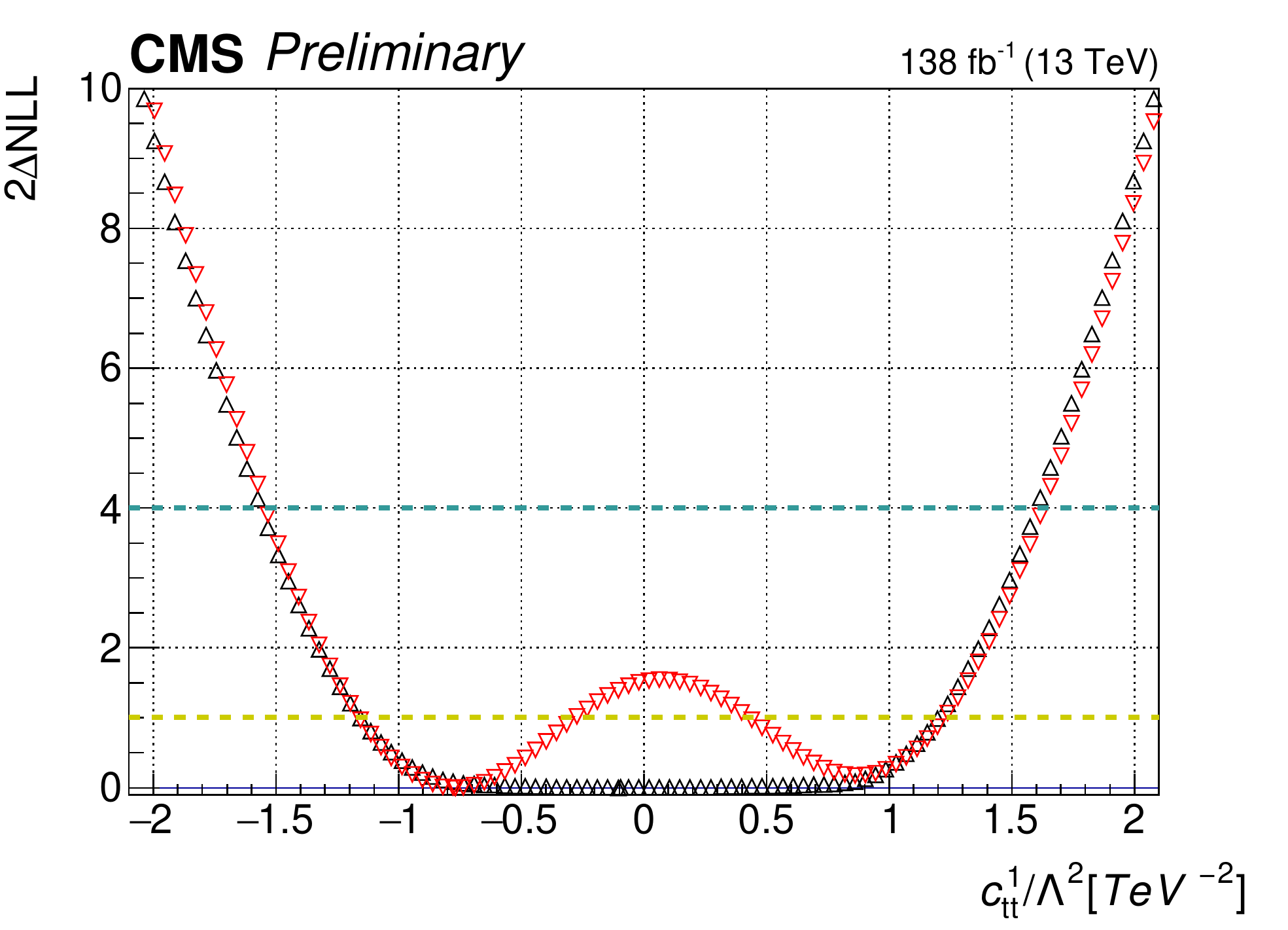} \\
		\includegraphics[width=\textwidth]{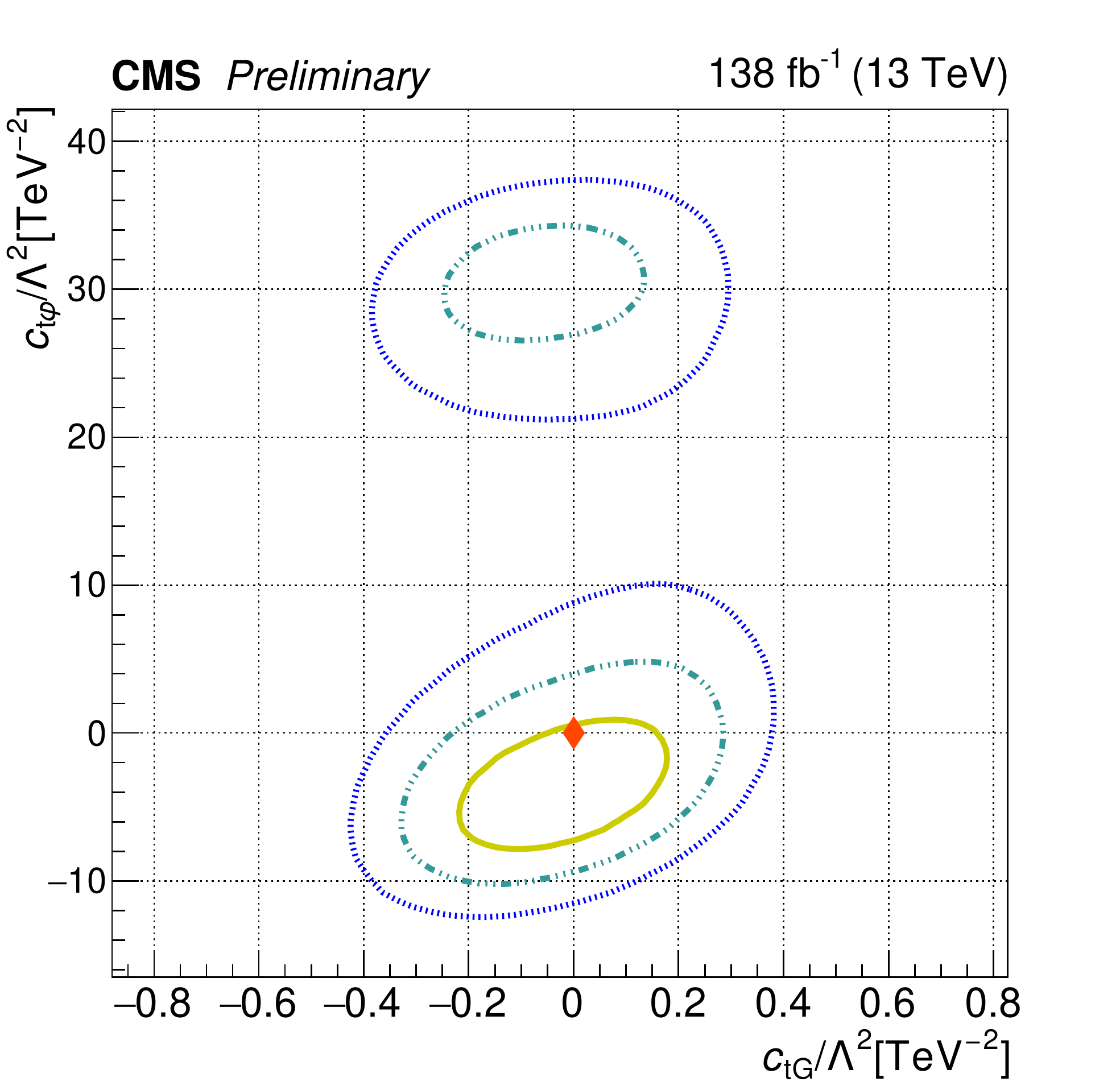} \\
	\end{minipage} 
	\caption{Left: confidence intervals on the WCs measuring all of them simultaneously (black) and assuming the rest are zero (red). Upper right: likelihod function as a function at \cttOne under the two hypotheses. Lower right: two-dimensional confidence regions as a function of \ctp and \ctG, profiling over the rest of WCs.}
	\label{fig:likelihoodfits}
\end{figure}

We extract limits on the WCs under two different sets of assumptions: one in which all WCs are considered as free parameters that are profiled in the fit and another one in which we extract limits on one or two WCs assuming the rest are zero. Figure~\ref{fig:likelihoodfits} shows the limits obtained on the 26 WCs we study along with the likelihood function as a function of \cttOne, both under the two different assumptions. Two-dimensional confidence regions are also shown as a function of \ctp and \ctG, profiling over the rest of WCs.

The different regions used in the analysis contribute differently to the sensitivity of each WCs. Table~\ref{tab:sensitivity} shows a simplified picture of the categories that dominate the sensitivity to each WC. Four fermion operators can be easily mapped to regions of the phase space that they contribute to: two heavy quark two lepton operators are predominantly constrained by the \threel off-$\PZ$, as they contribute dominantly to non-resonant \ttll and \ttlnu production; four heavy quark operators are constrained by the \twolss region with 3 \PQb-tagged jets, as it is enriched in \tttt events. There are two types of two heavy two light quark operators affecting $\ttlnu$ and $\PQt\ell\ell\PQb$ production, respectively, that are constrained by the \twolss and \threel on-Z regions. Two heavy quark with boson operators are constrained uniformly by all regions, except for those involving the top-Z coupling, which are predominantly constrained by the \threel on-Z region.

\begin{table}[!htb]
	\centering
	\caption{Summary of categories that provide leading contributions to the sensitivity for subsets of the WCs.}
	\begin{tabular}{ p{5.8cm} p{5.9cm} p{2.7cm}}
		\hline
		Grouping of WCs & WCs & Lead categories \\
		\hline
		 Two heavy two leptons  &  \cQla, \cQlM, \cQe, \ctl, \cte, \ctlS, \ctlT  &  \threel off-$\PZ$  \\ 
		Four heavy                             & \cQQOne, \cQtOne, \cQtEight, \cttOne          &  \twolss                       \\ 
		Two heavy two light   ``\ttlnu-like"   & \cQqOneOne, \cQqEightOne, \ctqOne, \ctqEight  &  \twolss                      \vspace{0.1cm} \\ 
		Two heavy two light   ``\tllq-like"    & \cQqOneThree, \cQqEightThree                  &  \threel on-$\PZ$             \vspace{0.1cm} \\ 
		Two heavy with bosons ``\ttll-like"    & \ctZ, \cpt, \cpQM                             &  \threel on-$\PZ$ and \twolss \\ 
		Two heavy with bosons ``\tZq-like"     & \cpQa, \cptb, \cbW                            &  \threel on-\PZ                          \vspace{0.1cm} \\ 
		Other two heavy with bosons & \ctG, \ctp, \ctW                              &  \threel and \twolss                           \\
		\hline
	\end{tabular}
	\label{tab:sensitivity}
\end{table}

Assuming particles in the UV theory have couplings to the SM of the order 1 (WCs$\sim$1), we can translate the limits shown in figure~\ref{fig:likelihoodfits} to limits on the energy scale of UV theories generating different kinds of interactions. Thus, these results allow us to exclude UV theories generating two heavy quark two lepton interactions up to 800 GeV to 1 TeV; theories generating two heavy quark with boson interactions up to 300 GeV to 1 TeV; theories generating four heavy quark interactions up to 700 GeV to 1 TeV, and theories generating two heavy two light quark interactions up to 1 to 3 TeV, all depending on the gauge structure of the UV theory.

\section{Summary and outlook}

We have performed a search for BSM physics in top quark production in association with additional leptons. These processes can be affected by several BSM scenarios that we parametrize in a model-agnostic way using the EFT framework. Assuming new physics couples naturally to the SM, we set limits on the presence of theories generating different kinds of interactions at energy scales up to between 300 GeV to 3 TeV.


\section*{References}

\begin{thebibliography}{99}
\bibitem{cms}  CMS Collaboration, JINST \textbf{3} (2008) S08004
\bibitem{TOP22006} CMS Collaboration, CMS-PAS-TOP-22-006
\bibitem{TOP19001} CMS Collaboration, JHEP \textbf{03} (2021) 095
\bibitem{Aguilar_Saavedra:2018ksv} Aguilar-Saavedra, J. A et al. \href{https://arxiv.org/abs/1802.07237}{arXiv:1802.07237}
\bibitem{HIG19008} CMS Collaboration, {\em Eur. Phys. J. C} {\bf 81} (2021), 378
\bibitem{HIG21006} CMS Collaboration, \href{http://arxiv.org/abs/2208.02686}{arXiv:2208.02686}. Accepted by JHEP
\bibitem{HIG18009}CMS Collaboration, {\em Phys. Rev. D} {\bf 99} (2019) 092005
\bibitem{TOP21011} CMS Collaboration, \href{http://arxiv.org/abs/2208.06485}{arXiv:2208.06485}. Accepted by JHEP
\bibitem{TOP18009} CMS Collaboration, JHEP \textbf{03} (2020) 056 
\bibitem{TOP20010} CMS Collaboration, JHEP \textbf{02} (2022) 107
\bibitem{TOP21005} CMS Collaboration,
 \href{http://arxiv.org/abs/2303.03864}{arXiv:2303.03864}. Submitted to PLB. 
\bibitem{TOP18006} CMS Collaboration, {\em Phys. Rev. D} {\bf 100} (2019) 072002 

\bibitem{reweighting} O. Mattelaer, {\em Eur. Phys. J. C} {\bf 76} (2016), 674



\end{thebibliography}
\end{document}